\documentclass[10pt, final, twocolumn, twoside, romanappendices]{IEEEtran}

\usepackage{graphicx,cite,acronym,amsmath,amssymb,amsfonts,color,subfigure,floatflt,stfloats,bm,textgreek,multirow}

\usepackage{epsfig,epstopdf,psfrag}

\DeclareGraphicsExtensions{.png,.jpg,.eps}

\acrodef{ESPAR}{electrically steerable passive array radiator}

\acrodef{SIM}{stacked intelligent metasurface}

\acrodef{DAC}{digital-to-analog converter}

\acrodef{DMA}{dynamic metasurface antenna}

\acrodef{SINR}{signal-to-interference noise ratio}

\acrodef{ESIT}{electromagnetic signal and information theory} 

\acrodef{ELAA}{extremely large antenna arrays} 

\acrodef{DSA}{dynamic scattering array}

\acrodef{ULA}{uniform linear array}

\acrodef{UCA}{uniform circolar array}

\acrodef{IIoT}{industrial Internet-of-things}

\acrodef{IT}{information theory}

\acrodef{SRE}{smart radio environment}

\acrodef{EMO}{electromagnetic object}

\acrodef{SVD}{singular value decomposition}

\acrodef{PSWF}{prolate spheroidal wave function}

\acrodef{CR}{channel response}

\acrodef{BS}{base station}

\acrodef{MS}{mobile station}

\acrodef{UE}{user equipment}

\acrodef{MIMO}{multiple-input multiple-output}

\acrodef{MISO}{multiple-input single-output}

\acrodef{RIS}{reconfigurable intelligent surface}

\acrodef{IRS}{intelligent reconfigurable surface}

\acrodef{LIS}{large intelligent surface}

\acrodef{MIS}{medium intelligent surface}

\acrodef{SIS}{small intelligent surface}

\acrodef{DoF}{degrees-of-freedom}

\acrodef{AF}{amplify \& forward}

\acrodef{DF}{detect \& forward}

\acrodef{JF}{just forward}

\acrodef{CSI}{channel state information}

\acrodef{RV}{random variable}

\acrodef{i.i.d.}{independent, identically distributed}

\acrodef{PSD}{power spectral density}

\acrodef{PDF}{probability distribution function}

\acrodef{CDF}{cumulative distribution function}

\acrodef{ch.f.}{characteristic function}

\acrodef{AWGN}{additive white Gaussian noise}

\acrodef{RSSI}{received signal strength indicator}

\acrodef{SNR}{signal-to-noise ratio}

\acrodef{LRT}{likelihood ratio test}

\acrodef{GLRT}{generalized likelihood ratio test}

\acrodef{GML}{generalized maximum likelihood}

\acrodef{LOS}{line-of-sight}

\acrodef{NLOS}{non-line-of-sight}

\acrodef{GDOP}{geometric dilution of precision}

\acrodef{GPS}{Global Positioning System}

\acrodef{FIM}{Fisher information matrix}

\acrodef{PEB}{position error bound}

\acrodef{WSN}{Wireless Sensor Network}

\acrodef{MAC}{medium access control}

\acrodef{RSS}{received signal strength}

\acrodef{RTT}{round-trip time}

\acrodef{MIMO}{multiple-input multiple-output}

\acrodef{MF}{matched filter}

\acrodef{ED}{energy detector}

\acrodef{ML}{maximum likelihood}

\acrodef{NL}{nonlinear}

\acrodef{MSE}{mean square error}

\acrodef{RMSE}{root mean square error}

\acrodef{ppm}{part-per-million}

\acrodef{PRP}{pulse repetition period}

\acrodef{ACK}{acknowledge}

\acrodef{UWB}{ultrawide bandwidth}

\acrodef{TNR}{threshold-to-noise ratio}

\acrodef{LOS}{line-of-sight}

\acrodef{LS}{least squares}

\acrodef{IR-UWB}{impulse radio UWB}

\acrodef{FCC}{Federal Communications Commission}

\acrodef{TH}{time-hopping}

\acrodef{PPM}{pulse position modulation}

\acrodef{PAM}{pulse amplitude modulation}

\acrodef{MUI}{multi-user interference}

\acrodef{PDP}{power delay profile}

\acrodef{PPP}{Poisson point process}

\acrodef{DS}{delay spread}

\acrodef{CED}{channel excess delay}

\acrodef{BPZF}{band-pass zonal filter}

\acrodef{SIR}{signal-to-interference ratio}

\acrodef{RFID}{radio frequency identification}

\acrodef{WPAN}{wireless personal area networks}

\acrodef{WWLB}{Weiss-Weinstein lower bound}

\acrodef{DP}{direct path}

\acrodef{MF}{matched filter}

\acrodef{MMSE}{minimum-mean-square-error}

\acrodef{SBS}{serial backward search}

\acrodef{NBI}{narrowband interference}

\acrodef{WBI}{wideband interference}

\acrodef{INR}{interference-to-noise ratio}

\acrodef{CIR}{channel impulse response}

\acrodef{ISI}{inter-symbol interference}

\acrodef{CPR}{channel pulse response}

\acrodef{LRT}{likelihood ratio test}

\acrodef{MUI}{multi-user interference}

\acrodef{EM}{electromagnetic}

\acrodef{CW}{continuous wave}

\acrodef{RF}{radiofrequency}

\acrodef{FCC}{Federal Communications Commission}

\acrodef{EIRP}{effective radiated isotropic power}

\acrodef{RCS}{radar cross section}

\acrodef{BAV}{balanced antipodal Vivaldi}

\acrodef{PRake}{partial Rake}

\acrodef{RTLS}{real time locating system}

\acrodef{CRB}{Cram\'{e}r-Rao bound}

\acrodef{ZZB}{Ziv-Zakai bound}

\acrodef{TOA}{time-of-arrival}

\acrodef{TOF}{time-of-flight}

\acrodef{WSN}{wireless sensor network}

\acrodef{MAC}{medium access control}

\acrodef{RSS}{received signal strength}

\acrodef{TDOA}{time difference-of-arrival}

\acrodef{RF}{radiofrequency}

\acrodef{RTT}{round-trip time}

\acrodef{AOA}{angle-of-arrival}

\acrodef{MF}{matched filter}

\acrodef{ED}{energy detector}

\acrodef{ML}{maximum likelihood}

\acrodef{MUR}{Multistatic radar}

\acrodef{HDSA}{high-definition situation-aware}

\acrodef{RRC}{root raised cosine}

\acrodef{OFDM}{orthogonal frequency division multiplexing}

\acrodef{IF}{intermediate frequency}

\acrodef{PHY}{physical layer}

\acrodef{S-V}{Saleh-Valenzuela}

\acrodef{UHF}{ultra-high frequency}

\acrodef{PR}{pseudo-random}

\acrodef{SoC}{System on Chip}

\acrodef{SoP}{System on Package}

\acrodef{SPMF}{Single-Path Matched Filter}

\acrodef{IMF}{Ideal Matched Filter}

\acrodef{SCR}{signal-to-clutter ratio}

\acrodef{BEP}{bit error probability}

\acrodef{BER}{bit error rate}

\acrodef{WSR}{wireless sensor radar}

\acrodef{HPBW}{half power beam width}

\acrodef{LEO}{localization error outage}

\acrodef{WSS}{wide-sense stationary}

\acrodef{TR}{time-reversal}

\acrodef{WSSUS}{WSS with uncorrelated scattering}

\acrodef{GP}{Gaussian process}

\acrodef{IMU}{inertial measurement unit}

\newcommand{\boldH} {{\bf{H}}}
\newcommand{\boldHopt} {{\bf{H}_{\mathrm{opt}}}}

\newcommand{\boldi} {{\bf{i}}}

\newcommand{\boldp} {{\bf{p}}}

\newcommand{\boldv} {{\bf{v}}}
\newcommand{\boldV} {{\bf{V}}}

\newcommand{\boldy} {{\bf{y}}}

\newcommand{\boldZ} {{\bf{Z}}}

\newcommand{\btheta} {\boldsymbol{\theta}}

\newcommand{\ctranspose}{{\mathrm{H}}}

\newcommand{\diag}[1]{{\rm diag} \left ( #1 \right )}

\newcommand{\Gr} {\mathbf{G}_{\text{R}}}

\newcommand{\Na} {N_{\text{A}}}

\newcommand{\Ns} {N_{\text{S}}}

\newcommand{\Prad} {P_{\text{rad}}}

\newcommand{\transpose}{{\mathrm{T}}}

\title{Frequency-selective Dynamic Scattering Arrays for Over-the-air EM Processing}

\author{
\IEEEauthorblockN{Davide~Dardari,~\IEEEmembership{Fellow,~IEEE}}
\IEEEcompsocitemizethanks{\IEEEcompsocthanksitem 
 D.~Dardari is with the 
   Dipartimento di Ingegneria dell'Energia Elettrica e dell'Informazione ``Guglielmo Marconi"  (DEI), WiLAB-CNIT, 
   University of Bologna, Cesena Campus, 
   Cesena (FC), Italy, (e-mail: davide.dardari@unibo.it). 
    }
}

\begin{document}

\maketitle

\begin{abstract}
  In this paper, we investigate frequency-selective \ac{DSA}, a versatile antenna structure capable of performing joint wave-based computing and radiation by transitioning signal processing tasks from the digital domain to the \ac{EM} domain.
The numerical results demonstrate the potential of \acp{DSA} to produce space-frequency superdirective responses with minimal usage of \ac{RF} chains, making it particularly attractive for future holographic \ac{MIMO} systems.
\end{abstract}

\section{Introduction}

The use of high-frequency bands in the millimeter wave and THz ranges, combined with the integration of antennas with a large number of elements, is driving current wireless technology toward seemingly insurmountable challenges in hardware complexity, latency, and power consumption. These issues present significant obstacles to the sustainability of future wireless networks.

One promising strategy for enhancing sustainability is to shift part of the signal processing directly to the \acf{EM} domain, a concept known as \ac{ESIT} \cite{DiRMig:24,BjoChaHeaMarMezSanRusCasJunDem:24,Dar:JS25}. 
This can be achieved by designing reconfigurable \ac{EM} environments \cite{Dar:J24} using advanced \ac{EM} metamaterial devices to perform basic processing functions \cite{Sil:14}, \acp{RIS}, or the recently introduced \acp{SIM} \cite{AnXuNgAleHuaYueHan:23}.
In particular, the study of \acp{SIM} remains in its early stages, and the key technological and modeling challenges are yet to be fully addressed. For example, the spacing between layers must span several wavelengths to ensure the validity of the cascade model currently in use. Additionally, power losses and signal distortion caused by the multiple layers and the bounding box require further investigation. A notable limitation of \acp{SIM} is that only the final layer radiates, restricting the characteristics of the generated \ac{EM} waves to those achievable with a conventional planar surface or array. Meanwhile, the hidden layers are primarily responsible for performing \ac{EM} transformations.

To address the aforementioned limitations, we investigate an \ac{EM} antenna structure, referred to as a \acf{DSA}, which offers significant flexibility by enabling advanced processing capabilities directly at the \ac{EM} level while radiating \cite{Dar:C24}. A \ac{DSA} comprises a limited number of active antenna elements, each connected to an \ac{RF} chain (input), and is surrounded by a cluster of many reconfigurable passive scatterers. These scatterers interact within the reactive near field, enabling \ac{EM} processing and radiation to occur jointly ``over the air".
A \ac{DSA} can be viewed as a generalization of reactively controlled arrays, initially introduced in \cite{Har:78}, and their evolution into \acp{ESPAR}, which utilize a single \ac{RF} chain \cite{BucJuaKamSib:20}.

In this paper, we extend the concept of \ac{DSA} to operate with wideband signals, thereby enabling space-frequency processing capabilities. Additionally, unlike previous works on \ac{DSA}, we eliminate the requirement for ideal loads and a perfectly reconfigurable matching network by introducing reconfigurable loads modeled as a realistic varactor diode. 
Building on the analytical expression of the frequency-dependent characteristics of the \ac{DSA} as a function of its configuration parameters, we formulate an optimization problem to design these parameters according to the desired space-frequency response.
Numerical results highlight the remarkable flexibility of the \ac{DSA} in realizing different beams each associated with a different port and frequency while offering significant advantages with respect to current holographic \ac{MIMO} solutions, such as reduced size and minimized number of \ac{RF} chains. For example, we demonstrate that the tight coupling of \ac{DSA} elements enables the realization of superdirectivity, regardless of the beam's direction - a capability that stands in contrast to standard arrays, where superdirectivity is typically limited to the end-fire direction.

\begin{figure}[t]
  \centering
  \includegraphics[width=1\columnwidth]{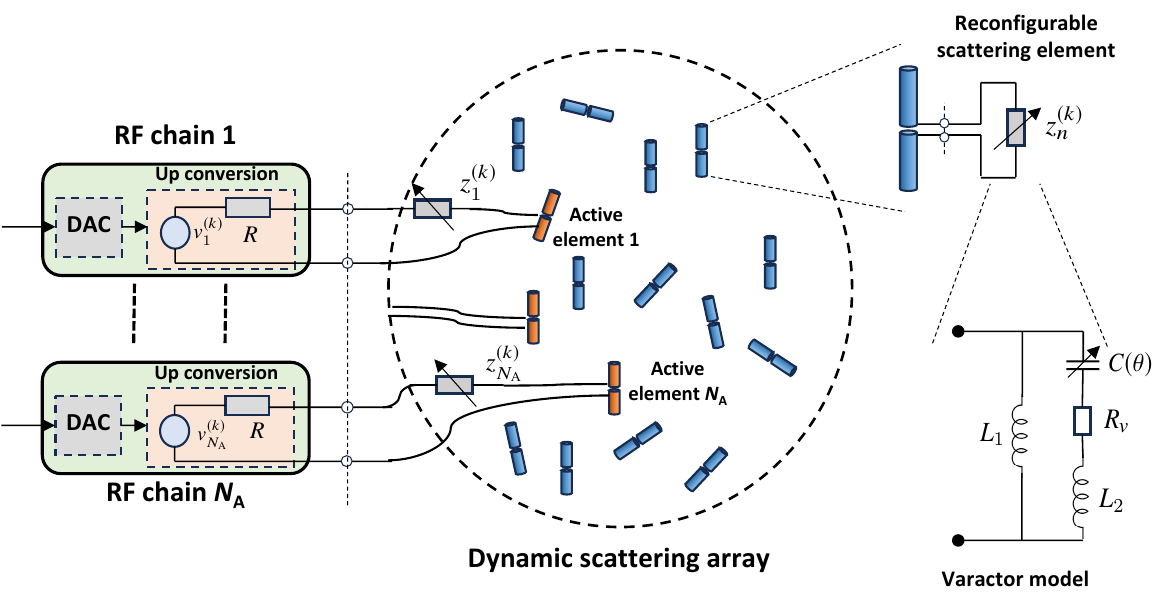}
  \caption{Schematic of the DSA.}
  \label{Fig:DSA}
\end{figure}

\section{Frequency-selective Dynamic Scattering Array Modelling}

With reference to Fig.~\ref{Fig:DSA}, a \ac{DSA} is composed of $N$ antenna elements of which $\Na$ are active and hence connected to an \ac{RF} chain, and $\Ns=N-\Na$ are passive scattering antenna elements \cite{Dar:C24}. 
 We consider a multicarrier system with total bandwidth $W$ using $K$ subcarriers at frequencies $f_k=f_0+(k-1) W/K$, for $k=1,2, \ldots K$. At the generic $k$th sucarrier, the $\Na$ \ac{RF} chains introduce open circuit voltages ${\boldv^{(k)}=\left [v_1^{(k)}, v_2^{(k)}, \ldots, v_{\Na}^{(k)}, 0, 0, \ldots, 0 \right]^{\transpose}}$ to the first $\Na$ elements that carry the information. 
The response of the \ac{DSA} can be changed by loading each antenna element with reconfigurable loads of impedances $z_n^{(k)}(\theta_n)$, with $n=1,2, \ldots N$, where $\theta_n$ is a parameter to be optimized, as it will be detailed later. Denote with $\btheta=\{ \theta_n\}_{n=1}^N$ the set of optimization parameters.  
Ideally, the loads should be designed to be only reactive to avoid power losses. 
At the $k$th subcarrier, the overall \ac{DSA} system can be modeled as an $N$-port network in which the following relationship between the set of currents $\boldi^{(k)}$ and open circuit voltages $\boldv^{(k)}$ is given by
 \begin{align}
 \left ( \boldZ_{\text{L}}^{(k)}(\btheta)  +\boldZ^{(k)}  \right ) \boldi^{(k)} =\boldv^{(k)}
\end{align}
where $\boldZ^{(k)} \in \mathbb{C}^{N \times N}$ is the impedance matrix of the  antenna structure and $\boldZ_{\text{L}}^{(k)}(\btheta)=\{ z_n^{(k)}(\theta_n) \}_{n=1}^N $ represents the set of reconfigurable loads' impedances. All the interactions between the elements of the \ac{DSA} are captured by the impedance matrix $\boldZ^{(k)}$, which does not depend on the reconfigurable loads, and relates the voltages and currents of the $N$ ports. For instance, with Hertzian dipoles or half-wave dipoles, it can be computed analytically \cite{BalB:16}.

Consider now $T$ test points located in positions $\boldp_t$, with $t=1,2, \ldots , T$, and that in each test point, a conventional receiving antenna is used to probe the signal with gain $\Gr$. 
As commonly done in the literature, we assume that the test points are located in the radiative region of the \ac{DSA} and the receiving antennas do not affect the transmitting \ac{DSA}. 
As an example, the set of $T$ antennas might represent a conventional receiving antenna array of a \ac{MIMO} system.  
The end-to-end relationship between the open circuit voltages at the $\Na$ ports and the useful component (i.e., without noise) of the received signal $\boldy^{(k)}=\left [y_1^{(k)},y_2^{(k)}, \ldots , y_T^{(k)} \right ]^{\transpose}$ at the test positions for subcarrier $k$  is 
\begin{align} \label{eq:by}
	\boldy^{(k)}=\boldH_c^{(k)} \, \boldi^{(k)}=\boldH_c^{(k)} \,   \left ( \boldZ_{\text{L}}^{(k)}(\btheta)  +\boldZ^{(k)}  \right )^{-1} \boldv^{(k)}   
\end{align}
where $\boldH_c^{(k)} \in \mathbb{C}^{T \times N}$ is the transimpedance matrix of the radio channel accounting for the propagation effects that can be easily computed analytically in free-space conditions.
The previous equation can be used as a general model for the \ac{DSA} in an optimization problem to determine the set of parameters $\btheta$ once the desired response $\boldy^{(k)}$ is fixed for each subcarrier.

Define the matrix $\boldV=\sqrt{4\, R} \, \diag{1,2, \ldots , \Na}  \in \mathbb{R}^{N \times \Na }$, being $R$ the internal resistance of the \ac{RF} chains. Indicating with $\boldi_n^{(k)}=\left ( \boldZ_{\text{L}}^{(k)}(\btheta)  +\boldZ^{(k)}  \right )^{-1} [\boldV]_{(:,n)}$ the current in the case only the $n$th \ac{RF} chain is excited at subcarrier $k$, where $[\boldV]_{(:,n)}$ represents the $n$th column of $\boldV$,  the corresponding radiated power results ${\Prad}_{,n}^{(k)}=\left ({\boldi_n^{(k)}}\right )^{\ctranspose} \Re\{\boldZ\} \, \boldi_n^{(k)} $, with $\Re\{x\}$ being the real part of $x$ and $(\cdot)^{\ctranspose}$ the conjugate transpose operator.  
The factor $\sqrt{4\, R}$ in matrix $\boldV$ has been introduced to normalize the power available at each \ac{RF} chain and subcarrier to $1\,$W.
Because of the losses and possible impedance mismatch between the \ac{RF} chains and the \ac{DSA} at the $\Na$ input ports, the efficiency of the \ac{DSA} is, in general, less than one, then ${\Prad}_{,n}^{(k)}\le 1\,$W.

In general, one may aim to solve the following  optimization problem for given objective end-to-end channel matrices  $\boldHopt^{(k)} \in \mathbb{C}^{T \times \Na}$, $k=1,2, \ldots, K$,   
\begin{equation} \label{eq:opt}
\hat{\btheta}=\arg \min_{\btheta} \sum_{k=1}^K \left\|\, \alpha^{(k)} \, \boldH_c^{(k)} \,   \left ( \boldZ_{\text{L}}^{(k)}(\btheta)  +\boldZ^{(k)}  \right )^{-1} \boldV -   \boldHopt^{(k)} \right\|^2_{\mathrm{F}}   
\end{equation}
where $\alpha^{(k)}$ is a normalization constant, and $\| \cdot \|_{\mathrm{F}}$ denotes the Fronebous norm.
The $t$th row of $\boldHopt^{(k)}$ represents the desired end-to-end channel response associated with the $n$th port and the $k$th subcarrier of the \ac{DSA}. It is worth noticing that with the same configuration $\hat{\btheta}$ of the \ac{DSA}, $\Na \times K$ different space-frequency responses are obtained simultaneously, each one associated with one specific subcarrier and the input port of one specific \ac{RF} chain.  
 
Regarding the reconfigurable load, we consider a varactor diode whose circuit model is reported in Fig. \ref{Fig:DSA} (right-bottom) as suggested in  \cite{SamRuiQinCha:20}. 
The corresponding impedance is
\begin{equation}
z_n^{(k)}(\theta_n)=\frac{j 2 \pi f_k L_1 (j 2 \pi f_k  L_2+1/(j 2 \pi f_k \, C(\theta_n))+R_v)}{j 2 \pi f_k (L_1+L_2)+1/(j 2 \pi f_k \, C(\theta_n))+R_v} 
\end{equation}
where $C(\theta)=C_{\text{min}}+(C_{\text{max}}-C_{\text{min}}) (\tan^{-1}(\theta)+\pi/2)/\pi$, being $C_{\text{min}}$ and $C_{\text{max}}$, respectively, the minimum and maximum capacitance of the varactor. The relationship between the optimization parameter $\theta$ and $C$ has been chosen so that $\theta$ is unbounded and the optimization problem in \eqref{eq:opt} becomes unconstrained, thus making its numerical solution computation faster.   
For convenience, we embed in $z_n^{(k)}(\theta_n)$, for $n=1,2, \ldots \Na$, also the internal resistance $R$ of the \ac{RF} chain. In the numerical results, the following parameters have been considered: $R_v=0.1\,$Ohm, $L_1=2.5\,$nH, $L_2=0.7 \,$nH, $C_{\text{min}}=0.47\,$pF, $C_{\text{max}}=2.35\,$pF \cite{SamRuiQinCha:20}.

\begin{figure}[t]
  \centering
  \includegraphics[width=0.8\columnwidth]{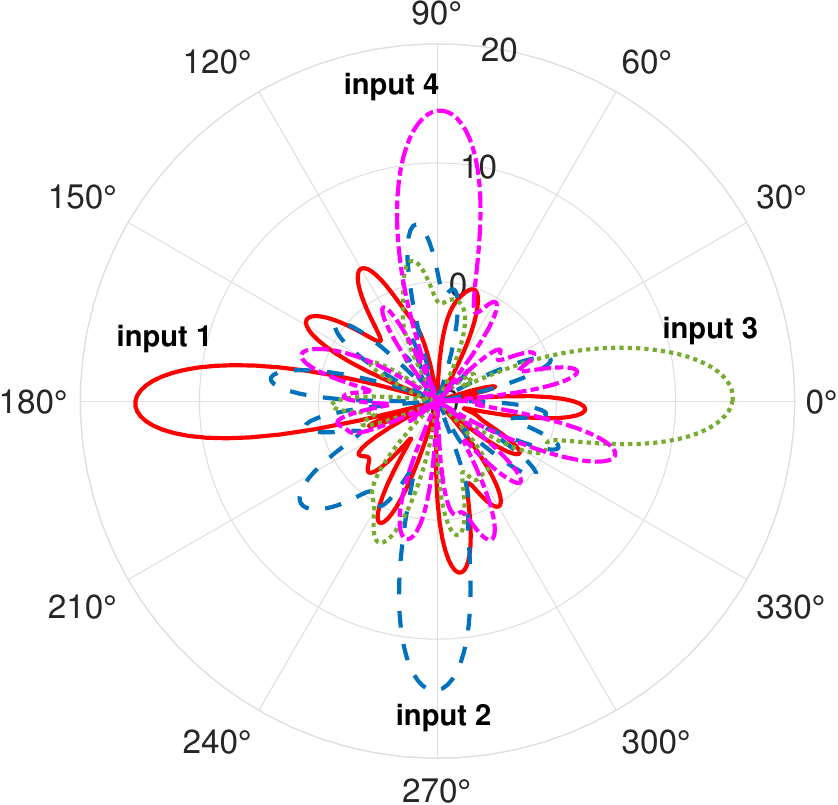}
  \caption{Single-carrier beam forming with $\Na=4$ \ac{RF} chains.}
  \label{Fig:PatternDSAMono}
\end{figure}

\begin{figure}[t]
  \centering
  \includegraphics[width=0.8\columnwidth]{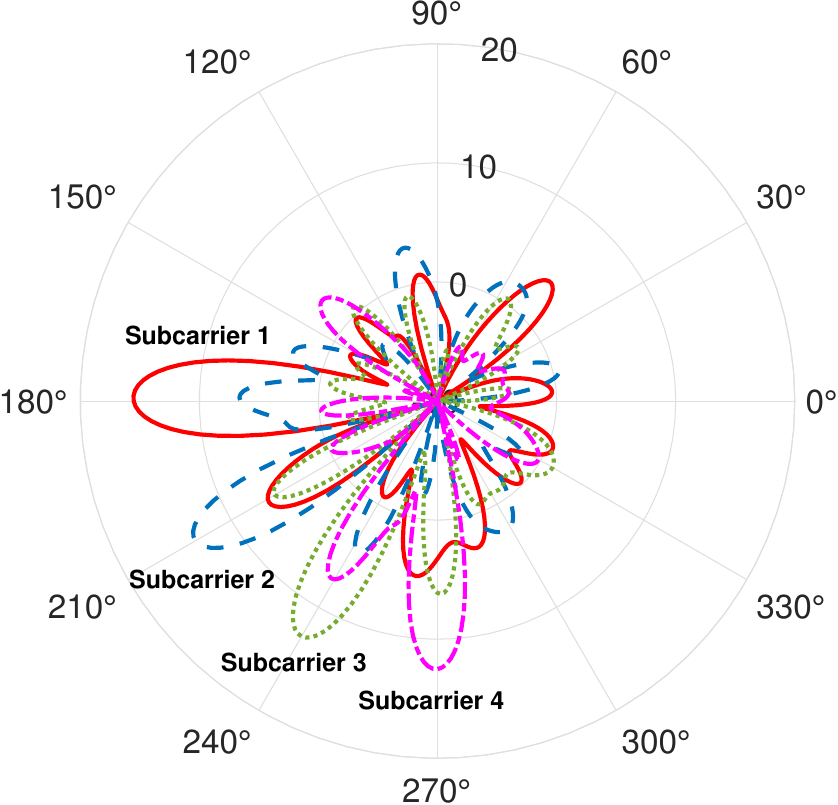}
  \caption{Multi-carrier beam forming with one \ac{RF} chain ($\Na=1$) and $K=4$ subcarriers.}
  \label{Fig:PatternDSAColor}
\end{figure}

\begin{figure}[t]
  \centering
  \includegraphics[width=0.9\columnwidth]{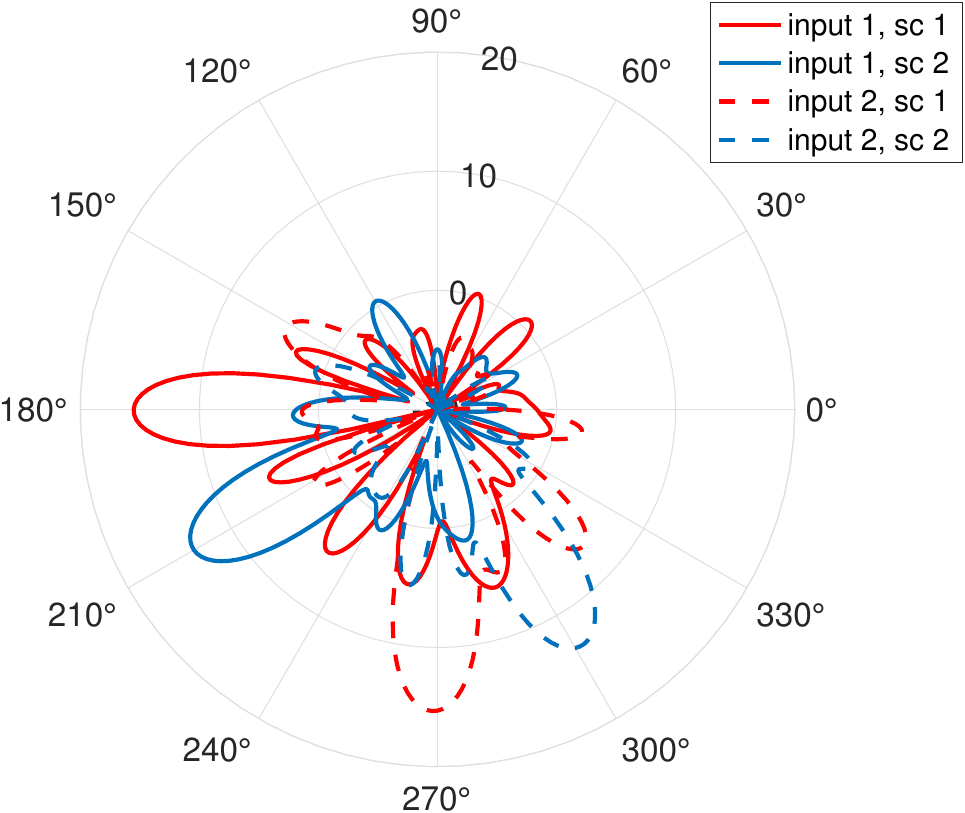}
  \caption{Joint angle-frequency beamforming  $\Na=2$ \ac{RF} chains and $K=2$ subcarriers.}
  \label{Fig:PatternHybrid}
\end{figure}

\section{Numerical Results}

The following parameters have been considered in the numerical evaluations if not otherwise specified: $f_0=2.4\,$GHz,  $R=75\,$Ohm, $\Gr=0\,$dB, $W=80\,$MHz. 
 Half-wave dipoles with vertical polarization ($z$-axis) have been considered for the analytical evaluation of the impedance matrix $\boldZ^{(k)}$ at each subcarrier's frequency \cite{BalB:16}. 
The scatterers of the \ac{DSA} were deployed according to $L=7$ concentric circles with an incremental radius of step  $\Delta=\lambda_0/4$ and separated of $\lambda/4$ along the circle and the $y$ axis corresponding to $\Ns=214$ scatterers, where $\lambda_0$ denotes the wavelength at $f_0$. Numerical investigations have put evidence that the spacing $\Delta=\lambda_0/4\,$ results in a better performance   \cite{Dar:C24}.  
This disk-like structure appears very appealing especially for its use in base stations or access points thanks to its circular symmetry. 
Obviously, other structures, even 3D, can be considered as well depending on the specific application. 

We first investigate the single-carrier beam forming capabilities of the \ac{DSA} using $\Na=4$ \ac{RF} chains (inputs). 
A uniform set of $T=108$  test points $\boldp_t=[d\sin \phi_t ,0 ,d \cos \phi_t]^\transpose$ on the $x-z$ plane a distance $d=100\,$m (far-field region) was considered, with $\phi_t=2 \pi t / T$.
The purpose is to associate a narrow beam radiation diagram to each \ac{RF} chain (i.e., data stream) steering towards a specific angle, respectively,  $180^{\circ}$, $270^{\circ}$, $0^{\circ}$, and $90^{\circ}$.  
This can be achieved  by setting   $[\boldHopt^{(1)}]_{(t_i,i)}=1$, for $i=1,2, \ldots \Na$ and zero otherwise, where $t_i$ denotes the index of the test point corresponding to the $i$th steering angle.   
The standard numerical tool based on the quasi-Newton method 
with $5000$ iterations has been utilized to minimize \eqref{eq:opt}.

In Fig.~\ref{Fig:PatternDSAMono}, the radiation diagrams for the 4 steering angles are shown.    
As it can be observed, the gain of the \ac{DSA} is independent of the angle and it is about $16-17\,$dB. In addition, limited back radiation is obtained without the need to insert a ground plane that would impede the steering in the opposite direction.
The \ac{DSA} exhibits superdirectivity capability with an additional gain of about $5\,$dB with respect to a conventional array with the same aperture of the  \ac{DSA}. 
Contrary to standard arrays, where superdirectivity is obtained only in the end-fire direction \cite{IvrNos:14}, superdirectivity is notably obtained in all steering directions.  Further examples of \ac{DSA} design for single-carrier systems to realize superdirective beamforming, and \ac{MIMO} precoding can be found in \cite{Dar:C24}. 
 
The capability of the \ac{DSA} to provide frequency-selective radiation diagrams can be observed in Fig.~\ref{Fig:PatternDSAColor}, where $K=4$ subcarriers where considered each of them associated with a different steering angle, respectively, $180^{\circ}$, $210^{\circ}$, $240^{\circ}$, and $270^{\circ}$, respectively. Compared to the single-carrier situation, the diagrams present slightly higher sidelobes indicating that managing wideband signals with the same configuration parameters of the \ac{DSA} is more challenging. 

Finally, the joint space-frequency performance of the \ac{DSA} is reported in Fig. \ref{Fig:PatternHybrid} obtained with $\Na=2$ \ac{RF} chains and   $K=2$ subcarriers. The target steering angles for each input/subcarrier (sc) combination are: $180^{\circ}$ (input 1, sc 1), $210^{\circ}$ (input 1, sc 2), $270^{\circ}$ (input 2, sc 1), and $300^{\circ}$ (input 2, sc 2). The plots show that with the same configuration $\btheta$, obtained from the minimization of \eqref{eq:opt}, the \ac{DSA} is capable of associating a dedicated superdirective beam to each input/subcarrier combination.

The gain and efficiency of the \ac{DSA} for the cases analyzed before, is reported in Table 1. In particular, $\eta_1$ and $\eta_2$ represent, respectively, the efficiency in case of a lossless \ac{DSA} caused by imperfect impedance matching at the input ports and the overall efficiency accounting also for the losses. As can be noticed, most of the efficiency decreases due to the losses indicating that a discrete impedance matching is typically achieved during the optimization process.

\begin{table}[tp]
\caption{}
\begin{center}
\begin{tabular}{|c|c|c|c|}
\hline
Fig. 2   \\
 \hline
 Configuration & Gain (dB) & $\eta_1$ & $\eta_2$ \\
 \hline
input 1   & 17 & 0.82 & 0.68 \\
input 2  & 16 & 0.84 & 0.68 \\ 
input 3 & 16.3 & 0.88 & 0.71 \\
input 4  & 16.3 & 0.82 & 0.64 \\
\hline
Fig. 3 \\
\hline
Configuration & Gain (dB) & $\eta_1$ & $\eta_2$ \\
 \hline
sc 1   & 16.7 & 0.97 & 0.83 \\
sc 2  & 15.1 & 0.97 & 0.69 \\ 
sc 3 & 15.3 & 0.98 & 0.56 \\
sc 4  & 15.7 & 0.92 & 0.48 \\
\hline
Fig. 4 \\
\hline
Configuration & Gain (dB) & $\eta_1$ & $\eta_2$ \\
 \hline
input 1, sc 1   & 16.6 & 0.95 & 0.78 \\
input 2, sc 2  & 15.9 & 0.95 & 0.60 \\ 
input 2, sc 1 & 16.5 & 0.94 & 0.77 \\
input 2, sc 2  & 15.6 & 0.97 & 0.59 \\
\hline
\end{tabular}
\end{center}
\label{Table:Cylinder}
\end{table}

\section{Conclusion}

In this paper, we have explored a frequency-selective \acf{DSA} as a promising technology designed to shift signal processing tasks from the digital domain to the \ac{EM} domain. This is achieved through the joint optimization of \ac{EM} processing and the radiation of a few active and many reconfigurable scattering elements interacting within the reactive near field.
The \ac{DSA} offers an interesting capability to manage the \ac{EM} field.
 This enables greater flexibility compared to traditional \ac{MIMO} and recent \ac{SIM} implementations. By minimizing the number of \ac{RF} chains, \ac{DSA} facilitates the development of energy-efficient, low-latency, and cost-effective holographic \ac{MIMO} systems, addressing the sustainability challenges of future wireless networks.

\section*{Acknowledgment}
This work was supported by the European Union under the Italian National Recovery and Resilience Plan (NRRP) of NextGeneration EU, partnership on ``Telecommunications of the Future" (PE00000001 - program ``RESTART"), and by the HORIZON-JU-SNS-2022-STREAM-B-01-03 6G-SHINE project (Grant Agreement No. 101095738).


\bibliographystyle{IEEEtran}
\bibliography{IEEEabrv,BiblioDD,MetaSurfaces,EMInformationTheory,IntelligentSurfaces,MassiveMIMO,MIMO,THzComm,EMTheory,WINS-Books,Vari}

\begin{thebibliography}{10}
\providecommand{\url}[1]{#1}
\csname url@samestyle\endcsname
\providecommand{\newblock}{\relax}
\providecommand{\bibinfo}[2]{#2}
\providecommand{\BIBentrySTDinterwordspacing}{\spaceskip=0pt\relax}
\providecommand{\BIBentryALTinterwordstretchfactor}{4}
\providecommand{\BIBentryALTinterwordspacing}{\spaceskip=\fontdimen2\font plus
\BIBentryALTinterwordstretchfactor\fontdimen3\font minus
  \fontdimen4\font\relax}
\providecommand{\BIBforeignlanguage}[2]{{%
\expandafter\ifx\csname l@#1\endcsname\relax
\typeout{** WARNING: IEEEtran.bst: No hyphenation pattern has been}%
\typeout{** loaded for the language `#1'. Using the pattern for}%
\typeout{** the default language instead.}%
\else
\language=\csname l@#1\endcsname
\fi
#2}}
\providecommand{\BIBdecl}{\relax}
\BIBdecl

\bibitem{DiRMig:24}
M.~D. Renzo and M.~D. Migliore, ``Electromagnetic signal and information
  theory,'' \emph{IEEE BITS the Information Theory Magazine}, pp. 1--13, 2024.

\bibitem{BjoChaHeaMarMezSanRusCasJunDem:24}
E.~{Bj{\"o}rnson}, C.-B. {Chae}, J.~{Heath}, Robert~W., T.~L. {Marzetta},
  A.~{Mezghani}, L.~{Sanguinetti}, F.~{Rusek}, M.~R. {Castellanos}, D.~{Jun},
  and {\"O}.~{Tugfe Demir}, ``{Towards {6G MIMO}: Massive Spatial Multiplexing,
  Dense Arrays, and Interplay Between Electromagnetics and Processing},''
  \emph{arXiv e-prints}, p. arXiv:2401.02844, Jan. 2024.

\bibitem{Dar:JS25}
\BIBentryALTinterwordspacing
D.~Dardari, G.~Torcolacci, G.~Pasolini, and N.~Decarli, ``An overview on
  over-the-air electromagnetic signal processing,'' 2024. [Online]. Available:
  \url{https://arxiv.org/abs/2412.14968}
\BIBentrySTDinterwordspacing

\bibitem{Dar:J24}
D.~Dardari, ``Reconfigurable electromagnetic environments: {A} general
  framework,'' \emph{IEEE Journal on Selected Areas in Communications},
  vol.~42, no.~6, pp. 1479--1493, June 2024.

\bibitem{Sil:14}
\BIBentryALTinterwordspacing
A.~Silva, F.~Monticone, G.~Castaldi, V.~Galdi, A.~Al{\`u}, and N.~Engheta,
  ``Performing mathematical operations with metamaterials,'' \emph{Science},
  vol. 343, no. 6167, pp. 160--163, 2014. [Online]. Available:
  \url{https://science.sciencemag.org/content/343/6167/160}
\BIBentrySTDinterwordspacing

\bibitem{AnXuNgAleHuaYueHan:23}
J.~An, C.~Xu, D.~W.~K. Ng, G.~C. Alexandropoulos, C.~Huang, C.~Yuen, and
  L.~Hanzo, ``Stacked intelligent metasurfaces for efficient holographic {MIMO}
  communications in {6G},'' \emph{IEEE Journal on Selected Areas in
  Communications}, vol.~41, no.~8, pp. 2380--2396, 2023.

\bibitem{Dar:C24}
D.~Dardari, ``{3D} electromagnetic signal processing,'' in \emph{Proc. Asilomar
  Conf. on Signals, Systems and Computers}, Oct 2024.

\bibitem{Har:78}
R.~Harrington, ``Reactively controlled directive arrays,'' \emph{IEEE Trans.
  Antennas Propag.}, vol.~26, no.~3, pp. 390--395, May 1978.

\bibitem{BucJuaKamSib:20}
J.~C. Bucheli~Garcia, M.~Kamoun, and A.~Sibille, ``Low-complexity adaptive
  spatial processing of {ESPAR} antenna systems,'' \emph{IEEE Trans. Wireless
  Commun."}, vol.~19, no.~6, pp. 3700--3711, Feb. 2020.

\bibitem{BalB:16}
C.~A. Balanis, \emph{Antenna Theory: {A}nalysis and {D}esign}.\hskip 1em plus
  0.5em minus 0.4em\relax New Jersey, USA: Wiley, 2016.

\bibitem{SamRuiQinCha:20}
S.~Abeywickrama, R.~Zhang, Q.~Wu, and C.~Yuen, ``Intelligent reflecting
  surface: Practical phase shift model and beamforming optimization,''
  \emph{IEEE Transactions on Communications}, vol.~68, no.~9, pp. 5849--5863,
  2020.

\bibitem{IvrNos:14}
M.~T. Ivrlac and J.~A. Nossek, ``The multiport communication theory,''
  \emph{IEEE Circuits and Systems Magazine}, vol.~14, no.~3, pp. 27--44, 2014.

\end{thebibliography}

\end{document}